\documentclass[10pt,twocolumn,aps,prl,superscriptaddress,showpacs,floatfix]{revtex4-1}

\usepackage{amsmath}
\usepackage{amssymb}
\usepackage{braket}
\usepackage{graphicx}
\usepackage{verbatim}
\usepackage{color}
\usepackage{amsthm}

\begin{document}

\title{Shortcuts to nonabelian braiding}

\author{Torsten Karzig}
\affiliation{Institute of Quantum Information and Matter, Department of Physics, California Institute of Technology, Pasadena, California 91125, USA}

\author{Falko Pientka}
\affiliation{\mbox{Dahlem Center for Complex Quantum Systems and Fachbereich Physik, Freie Universit\"at Berlin, 14195 Berlin, Germany}}

\author{Gil Refael}
\affiliation{Institute of Quantum Information and Matter, Department of Physics, California Institute of Technology, Pasadena, California 91125, USA}

\author{Felix von Oppen}
\affiliation{\mbox{Dahlem Center for Complex Quantum Systems and Fachbereich Physik, Freie Universit\"at Berlin, 14195 Berlin, Germany}}

\begin{abstract}
Topological quantum information processing relies on adiabatic braiding of nonabelian quasiparticles. Performing the braiding operations in finite time introduces transitions out of the ground-state manifold and deviations from the nonabelian Berry phase. We show that these errors can be eliminated by suitably designed counterdiabatic correction terms in the Hamiltonian. We implement the resulting shortcuts to adiabaticity for simple protocols of nonabelian braiding and show that the error suppression can be substantial even for approximate realizations of the counterdiabatic terms. 
\end{abstract}


\maketitle

{\em Introduction.---}It is envisaged that the information processing of topological quantum computers relies on adiabatic braiding of nonabelian quasiparticles \cite{kitaev03,freedman98,review,review2,review3}. Exchanging two nonabelions does not leave the quantum state unchanged, possibly up to a sign (as for fermions or bosons) or phase (as for abelian anyons) factor, but rather effects a unitary rotation in a degenerate subspace of ground states. The ground-state degeneracy grows exponentially with the number of nonabelian quasiparticles, and quantum information processing corresponds to manipulating the system's ground state by braiding operations. Majorana bound states in topological superconducting phases constitute the simplest example of such nonabelions \cite{1DwiresKitaev}, and there has been considerable experimental effort towards realizing a possible hardware \cite{mourik12,das12,churchill13,rokhinson12,lund,harlingen,yazdani}, following a series of theoretical proposals \cite{read00,fu08,fu09,sau10,alicea10,lutchyn10,oreg10,bernevig,pientka13}. 

Topological quantum information processing is immune to local sources of decoherence when braiding is performed adiabatically \cite{kitaev03}. Quite generally, adiabaticity is protected by the gap of the underlying topological phase. Here we want to ask the question whether it is possible to realize the {\em exact} adiabatic quantum dynamics of the braiding operation, albeit in a {\em finite} time interval. There are obvious motivations why this would be desirable: First, any topological quantum computer would operate at a finite clock speed which necessarily entails possibly small, but nonzero errors. Second, a topological quantum computer would presumably have to operate faster than parasitic decoherence processes such as quasiparticle poisoning or deviations from perfect ground state degeneracy originating in the finite spatial extent of the Majorana quasiparticles. In both cases, such a scheme could then be used to offset the incurred errors -- enabling longer computations or higher clock speeds. 

Demirplak and Rice \cite{demirplak} as well as Berry \cite{berry} introduced a protocol that emulates the adiabatic dynamics of any {\em nondegenerate} Hamiltonian $H_0(t)$ as the exact quantum dynamics in finite time. This scheme is known alternately as transitionless quantum driving or shortcut to adiabaticity. The prize that comes with the shortcut is that the adiabatic quantum dynamics of $H_0(t)$ is generated by a Hamiltonian $H(t)$, which differs from $H_0$ by counterdiabatic terms $H_1(t)$. This shortcut to adiabaticity does not apply directly to the adiabatic braiding of nonabelian quasiparticles because of the associated ground-state degeneracy. Here, we first generalize this scheme to systems with degenerate manifolds of states where adiabatic dynamics generates nonabelian Berry phases. Then, we apply this generalized shortcut to nonabelian statistics, using a simple model for braiding of Majorana bound states. Within this model, the braiding of Majorana zero modes is based on judiciously chosen temporal variations of the couplings between a number of Majorana end states. We find that shortcuts to nonabelian braiding can be implemented by introducing a small number of additional local couplings. 

{\em Shortcuts to adiabaticity for degenerate systems.---}The exact quantum dynamics of a Hamiltonian $H(t)$ is generated by the corresponding time-evolution operator ${\mathcal U}(t)$ which satisfies the Schr\"odinger equation
\begin{equation}
    i\partial_{t}{\mathcal{U}}(t)={H}(t){\mathcal{U}}(t).
\end{equation}
Thus, we can give an explicit expression for the Hamiltonian $H(t)$ generating any prescribed quantum dynamics ${\mathcal{U}}(t)$,
\begin{equation}
    {H}(t) = i[\partial_{t}{\mathcal{U}}]{\mathcal{U}}^\dagger.
\label{huu}
\end{equation}
The shortcut to adiabaticity \cite{demirplak,berry} follows by inserting into this expression the adiabatic time-evolution operator 
\begin{equation}
  {\mathcal{U}}(t)=\sum_n{\rm e}^{-i\int_{0}^{t}{\rm d}t'E_{n}(t')+i\gamma_n(t)}|\psi_{n}(t)\rangle\langle\psi_{n}(0)|
\label{eq:time_evolution}
\end{equation}
for the Hamiltonian $H_0(t)$, with instantaneous eigenvalues $E_n(t)$, instantaneous eigenstates $|\psi_n(t)\rangle$, and Berry phase $\gamma_n(t) =i\int_{0}^{t}{\rm d}t'\langle\psi_{n}(t')|\partial_{t'}\psi_{n}(t')\rangle$. One finds that $H(t)=H_0(t)+H_1(t)$ with the so-called counterdiabatic terms \cite{demirplak,berry} 
\begin{equation}
    {H}_{1}(t)=i\sum_{n}\left(|\partial_{t}\psi_{n}\rangle\langle\psi_{n}|-   
    |\psi_{n}\rangle\langle\psi_{n}|\partial_{t}\psi_{n}\rangle\langle\psi_{n}|\right).
  \label{eq:abelian_counter}
\end{equation}
Such shortcuts to adiabaticity have recently been implemented experimentally for effective two-level systems arising in trapped Bose-Einstein condensates \cite{fazio} and for the electron spin of a single nitrogen vacancy center \cite{suter}.

Following Wilczek and Zee \cite{zee}, we now consider a Hamiltonian $H_0(t)$ whose instantaneous spectrum defined through
\begin{equation}
   H_0(t) |\psi^n_\alpha(t)\rangle = E_n(t) |\psi^n_\alpha(t)\rangle
\label{instant}
\end{equation}
includes one or more sets of states $|\psi^n_\alpha(t)\rangle$ which remain degenerate for all $t$. Here, $\alpha=1,\ldots,d_n$ labels the states within the degenerate subspace $n$ of multiplicity $d_n$. 

We first define $|\eta^n_\alpha(t)\rangle$ as the adiabatic solution of the time-dependent Schr\"odinger equation
\begin{equation}
    i\partial_t |\eta^n_\alpha(t)\rangle = H_0(t) |\eta^n_\alpha(t)\rangle
\end{equation}
with initial condition $|\eta^n_\alpha(0)\rangle = |\psi^n_\alpha(0)\rangle$. In the adiabatic limit, the time-evolved state need not remain parallel to $|\psi^n_\alpha(t)\rangle$ but will in general be a linear combination of all basis states within the degenerate subspace,
\begin{equation}
   |\eta^n_\alpha(t)\rangle = \sum_\beta U^n_{\alpha\beta} |\psi^n_\beta(t)\rangle.
\end{equation}
Inserting this expansion into the time-dependent Schr\"odinger equation yields an equation for the coefficient matrices $U^n$, 
\begin{equation}
  i\partial_t U^n = U^n(A^n - E_n {\bf 1}),
  \label{schroedi}
\end{equation}
where $A^n_{\alpha\beta}=i\langle \psi^n_\beta|\partial_t \psi^n_\alpha\rangle$ denotes the nonabelian Berry connection  \cite{zee}. This is solved by
\begin{equation}
  U^n(t) = e^{-i\int_0^t dt^\prime E_n(t^\prime)}{\tilde T} e^{i\int_0^t dt^\prime A^n(t^\prime)}
\end{equation}
in terms of time-ordered exponentials.

The adiabatic time evolution of the Hamiltonian $H_0(t)$ follows the time-evolution operator
\begin{equation}
  {\cal U}(t) = \sum_{n,\alpha} |\eta^n_\alpha(t)\rangle\langle \eta^n_\alpha(0)| 
     =  \sum_{n,\alpha\beta} U^n_{\alpha\beta}|\psi^n_\beta(t)\rangle\langle \psi^n_\alpha(0)|.
\label{una}
\end{equation}
Now we use Eq.\ (\ref{huu}) to derive the Hamiltonian $H(t)$ for which this is the {\em exact} time-evolution operator. Inserting Eq.\ (\ref{una}) into (\ref{huu}), we obtain the shortcut to adiabaticity (all quantities evaluated at time $t$)
\begin{eqnarray}
   H &=& i\sum_n\sum_{\alpha\beta} \left\{ [(U^n)^\dagger {\dot U}^n]_{\beta\alpha} |\psi^n_\alpha\rangle\langle\psi^n_\beta| \right.\nonumber\\
   && \,\,\,\,\,\,\,\,\,\,\,\,\,\,\,\,\,\,\,\,\,\,\,\,\,\,
    \left. + [(U^n)^\dagger {U}^n]_{\beta\alpha}|\partial_t\psi^n_\alpha\rangle\langle\psi^n_\beta|\right\}
\end{eqnarray}
The second term in $H$ simplifies due to unitarity of $U^n$, $(U^n)^\dagger U^n={\bf 1}$. Combining unitarity and Eq.\ (\ref{schroedi}), we also have
$i(U^n)^\dagger {\dot U}^n = (E_n {\bf 1} - A^n)$ which simplifies the first term. With these identities, we readily find $H(t) = H_0(t) + H_1(t)$ with 
\begin{equation}
    {H}_{1}={\rm i}\sum_{n}\left[\sum_{\alpha}|\partial_{t}\psi^{n}_\alpha\rangle\langle\psi^{n}_\alpha|- \sum_{\alpha\beta}  
    |\psi^{n}_\alpha\rangle\langle\psi^{n}_\alpha|\partial_{t}\psi^{n}_\beta\rangle\langle\psi^{n}_\beta|\right].
  \label{nonabelian_counter}
\end{equation}
These counterdiabatic terms generalize the shortcut to adiabaticity to systems with degenerate spectra and nonabelian Berry connections. 

{\em Majorana systems.---}In view of topological quantum information processing, we specifically consider the counterdiabatic terms for a Bogoliubov--de Gennes Hamiltonian in Majorana representation,
\begin{align}
 H_0=i\sum_{n\alpha}\epsilon_n\gamma_{n,2\alpha-1}\gamma_{n,2\alpha}.\label{dia-M}
\end{align}
Here, both $\epsilon_n$ and the $\gamma_{n,\alpha}$ are explicitly time dependent and associated with the instantaneous Hamiltonian. The instantaneous many-body spectrum of $H_0$ contains degeneracies whenever an eigenenergy $\epsilon_n$ vanishes or when one or several nonzero $\epsilon_n$ are degenerate.  The Majorana eigenmodes associated with $\epsilon_n$ are denoted  by $\gamma_{n,\alpha}$ where $\alpha$ takes on $2N$ values for an $N$-fold degenerate energy $\epsilon_n$. The counterdiabatic terms $H_1$ guarantee that the time evolution generated by the full Hamiltonian $H_0+H_1$ does not take the Majorana eigenmodes $\gamma_{n,\alpha}$ out of the subspace $n$. At the same time, $H_1$ should not alter the time evolution within these subspaces. In the supplementary material \cite{supp} we show that these conditions yield
\begin{equation}
	H_1 = \frac{i}{4}\sum_{n\alpha} \dot{\gamma}_{n,\alpha}\gamma_{n,\alpha}-\frac{i}{8} \sum_{n,\alpha\beta}\gamma_{n,\alpha}\{\gamma_{n,\alpha},\dot{\gamma}_{n,\beta}\}\gamma_{n,\beta}.\label{fh1}
\end{equation}
This result complements the counterdiabatic terms in first quantization in Eq.~(\ref{nonabelian_counter}). 

\begin{figure}[t]
\includegraphics[width=.48\textwidth]{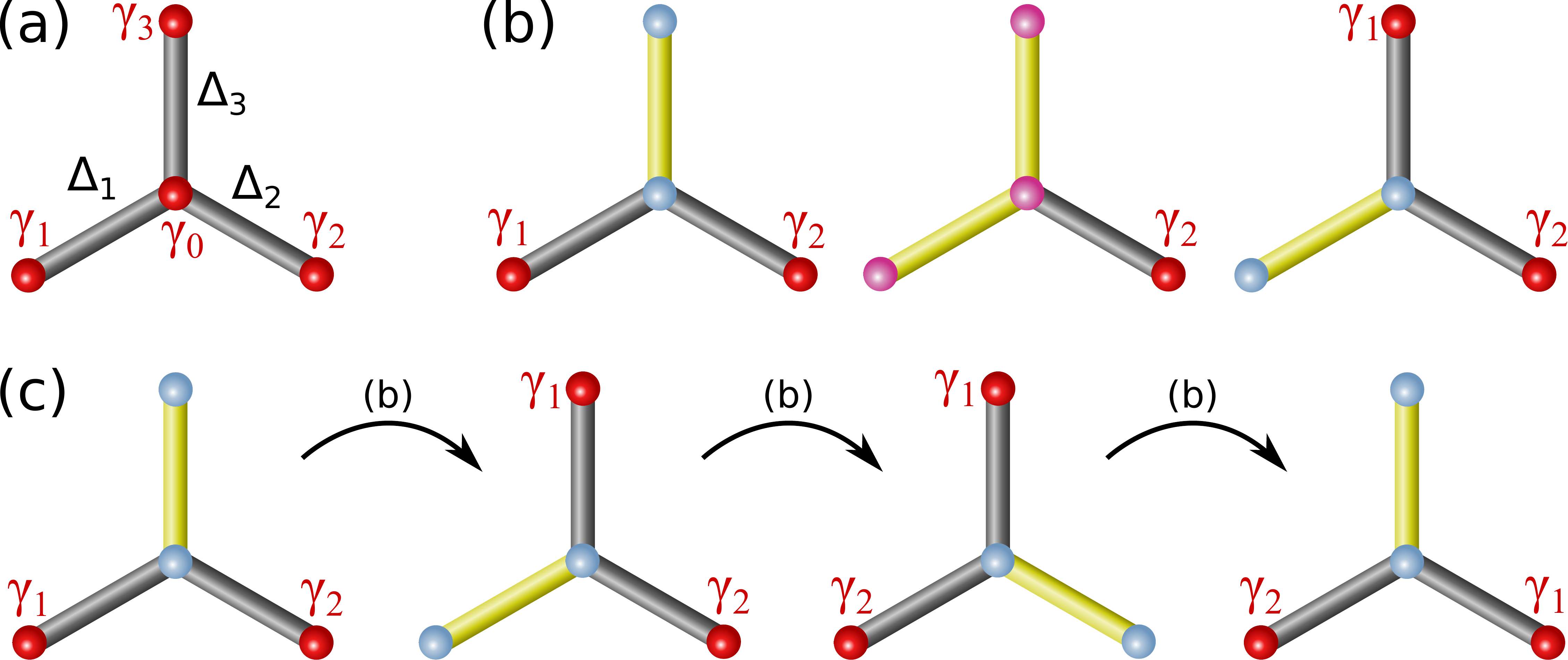}
\caption{(a) Y-junction with central Majorana $\gamma_0$ and three outer Majoranas $\gamma_j$ ($j=1,2,3$). The outer Majoranas are coupled to the inner Majoranas with strength $\Delta_j$. (b) Basic step of the braiding procedure, moving a zero-energy Majorana from the end of wire 1 to the end of wire 3 by varying the $\Delta_j$. Dark (light) wires indicate zero (nonzero) couplings $\Delta_j$. Dark red circles correspond to zero-energy Majoranas, light blue circles indicate Majoranas acquiring a finite energy by coupling. In the intermediate step, the zero-energy Majorana is delocalized over the three Majoranas along the light wires. (c) Three steps as described in (b) result in braiding the zero-energy Majoranas $\gamma_1$ and $\gamma_2$.}
\label{fig1}
\end{figure}

{\em Application to nonabelian braiding.---}A minimal model for nonabelian braiding starts from a Y-junction of three one-dimensional topological superconductors, labeled wire 1, 2, and 3 \cite{alicea,sau,heck}, as illustrated in Fig.\ \ref{fig1}(a). If all three arms are in the topological phase, there are four Majorana bound states in this system. Three of these are located at the outer ends of the three wires, with Bogoliubov operators labeled $\gamma_j$ for wire $j$, and a fourth Majorana mode $\gamma_0$ is located at the junction of the three wires. As long as the three arms have a finite length, these outer Majorana bound states hybridize with the central Majorana and the system is described by the Hamiltonian
\begin{equation}
    H_0 = i\sum_{\alpha=1}^3 \Delta_{\alpha}\gamma_0\gamma_{\alpha}
  \label{h0}
\end{equation} 
This Hamiltonian couples the central Majorana $\gamma_0$ to a linear combination of the outer three Majoranas. We can thus readily bring it to the form of Eq.\ (\ref{dia-M}),
\begin{equation}
H_0=i h_{\Delta}\gamma_0\gamma_{\Delta},
  \label{h0-dia}
\end{equation} 
with $\gamma_{\Delta}=(1/h_{\Delta})\sum_{\alpha=1}^3\Delta_{\alpha}\gamma_{\alpha}$ and 
$h_{\Delta}=[\Delta_1^2 + \Delta_2^2 +\Delta_3^2]^{1/2}$. For any choice of the couplings $\Delta_j$, there are also two linearly independent combinations of the outer Majoranas which do not appear in the Hamiltonian and thus remain true {\em zero-energy} Majoranas. Due to these zero-energy modes, the two eigenvalues of $H_0$ are each doubly degenerate. Specifically, when just one of the couplings $\Delta_j$ is nonzero, these two zero-energy Majoranas can be identified with the Majoranas located at the ends of those wires with zero coupling. 

We assume that we can change the couplings $\Delta_j$ as a function of time. We can now imagine the following braiding procedure \cite{alicea,heck}. Initially, only $\Delta_3$ is nonzero. Then, $\gamma_1$ and $\gamma_2$ are zero-energy Majoranas. In a first step, we move a zero-energy Majorana from the end of wire 1 to the end of wire 3, without involving the zero-energy Majorana $\gamma_2$ as shown in Fig.\ \ref{fig1}(b). To this end, first increase $\Delta_1$ to a finite value. The zero-energy Majorana originally located at the end of wire 1 is now delocalized and a linear combination of $\gamma_1$ and $\gamma_3$. We then localize the Majorana zero mode at the end of wire 3 by reducing $\Delta_3$ down to zero, leaving only $\Delta_1$ nonzero. The braiding process is completed by two completely analogous moves (see Fig.\ \ref{fig1}(c)): We first move the zero-energy Majorana from the end of wire 2 to the end of wire 1, and finally the zero-energy Majorana from wire 3 to wire 2. The combined effect of this procedure is to exchange the initial zero-energy Majoranas at the ends of wires 1 and 2. One can check easily \cite{heck} that the change of the state of the system under this adiabatic exchange is described by the familiar braiding matrix $U_{12} = \exp(i\pi \gamma_1\gamma_2/4)$. 

\begin{figure}[t]
\includegraphics[width=.48\textwidth]{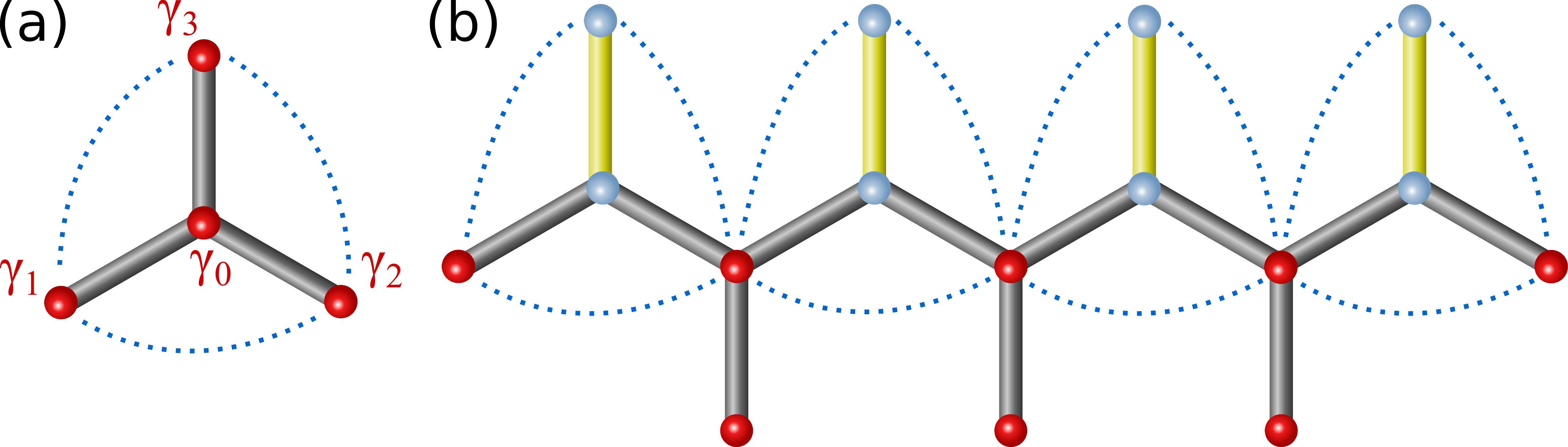}
\caption{(a) Minimal implementation required for braiding with shortcut protocol. The additional couplings needed for the shortcut protocol are shown in blue. (b) Wire network with many Majoranas allowing for pairwise exchanges of neighboring Majoranas including shortcut protocol. Implementing the shortcut merely requires the addition of local couplings within the network.}
\label{fig2}
\end{figure}

When performing this exchange operation over a finite time interval, there will be corrections to the adiabatic time evolution. We can now apply one of the nonabelian shortcut formulas in Eqs.\ (\ref{nonabelian_counter}) or (\ref{fh1}). As shown in the supplementary material \cite{supp}, we obtain
\begin{equation}
   H_1 =\frac{i}{2}\dot\gamma_{\Delta}\gamma_{\Delta}= \frac{i}{2h_{\Delta}^2}\sum_{\alpha<\beta} (\Delta_{\beta}\dot\Delta_{\alpha}-\Delta_{\alpha}\dot\Delta_{\beta})\gamma_{\alpha}\gamma_{\beta}. 
\label{shortcut2braiding}
\end{equation}
Thus, the shortcut is based on additional couplings between the outer Majoranas, while the adiabatic braiding protocol only uses couplings between the central and the outer Majoranas, see Fig.\ \ref{fig2}(a). Specifically, during the basic step of moving a zero-energy Majorana from the end of wire $i$ to wire $j$, only the couplings $\Delta_i$ and $\Delta_j$ are nonzero. According to Eq.\ (\ref{shortcut2braiding}), performing this step accurately in finite time merely requires the additional coupling between $\gamma_i$ and $\gamma_j$. 

{\em Practical implementation.---}There has been considerable work on how to implement braiding based on one-dimensional superconducting phases \cite{alicea,heck,sau,flensberg,halperin,hyart}. The couplings of the Majoranas can, e.g., be varied by changing the length of the intervening topological section. However, this may not be easily compatible with the geometric constraints imposed by the shortcut protocol, cf.\ Fig.\ \ref{fig2}(a). A better approach may be to vary the magnitude of the topological gap. Both methods control the overlap of the Majorana end states and hence their coupling. Physically, this can be achieved, say in quantum-wire based realizations, by changing the chemical potential by means of a gate electrode \cite{alicea} or a supercurrent in the adjacent $s$-wave superconductor \cite{romito}. 

More controlled variations of the Majorana couplings may be possible by exploiting charging effects \cite{heck} or by quantum dots \cite{flensberg}. For simplicity, assume that the quantum dot has a single level which is tunnel coupled to the ends of two topological wires with their Majorana end states. When the dot level is far from the Fermi energy, there is essentially no coupling between the adjacent Majoranas. Conversely, when the dot level is close to the Fermi energy, the Majoranas become strongly coupled. This approach modifies the coupling of the Majoranas by conventional gate control of a quantum-dot level and is also compatible with the geometric constraints of the shortcut protocol. 

So far, we have focused on the exchange of two Majoranas within the minimal setting of a Y-junction. Of course, one can readily imagine a scheme in which there is an entire keyboard of Majoranas and any two neighboring Majoranas can be readily braided. Importantly, amending this scheme to implement the counterdiabatic terms merely requires additional {\em local} couplings as shown in Fig.\ \ref{fig2}(b).

{\em Robustness.---}The manipulation of the quantum state is independent of the precise braiding path as long as the exchange is performed adiabatically. In contrast, the diabatic corrections are sensitive to the details of the braiding protocol. Consequently, the counterdiabatic terms (\ref{shortcut2braiding}) are not topologically protected, depend on the specifics of the braiding path, and for full effect, have to be implemented exactly for a given $H_0(t)$. 

However, we find that one can reach substantial reductions in the diabatic errors even when the shortcut protocol is implemented only with reasonable accuracy. We have computed the diabatic errors numerically, both for the bare braiding protocols and for approximate implementations of the counterdiabatic terms. Specifically, we consider the diabatic errors for 
\begin{equation}
  H_\lambda(t) = H_0(t)+\lambda H_1(t).\label{lambda}
\end{equation}
For $\lambda=1$, the counterdiabatic terms exactly compensate the diabatic corrections for any duration of the braiding protocol. As approximate implementations of the counterdiabatic terms, we consider relative errors of 10\% ($\lambda=0.9$) and 30\% ($\lambda=0.7$). We compute both the transition probability out of the degenerate subspace and the accumulated deviation from the adiabatic Berry phase.

Implementing the basic step [shown in Fig.\ \ref{fig1}(b)] of the braiding protocol in Fig.\ \ref{fig1}(c) by $\Delta_1(t)=\Delta\sin\varphi(t)$ and $\Delta_3(t)=\Delta\cos\varphi(t)$, with $\varphi(t)$ increasing from 0 to $\pi/2$, both the transition probability and the phase error exhibit a power-law dependence on the protocol duration $T$. The power law depends on the specific choice for $\varphi(t)$. Choosing the latter such that the derivative vanishes at the end points yields a $T^{-4}$ dependence. Corresponding numerical results are included with the supplementary material \cite{supp}. Interestingly, we find similar results for the protocol given in Ref.\ \cite{heck}, in which one initially increases $\Delta_1$, leaving $\Delta_3$ constant, and then reduces $\Delta_3$ to zero in a second step \cite{supp}. 
 
\begin{figure}[t]
\includegraphics[width=.48\textwidth]{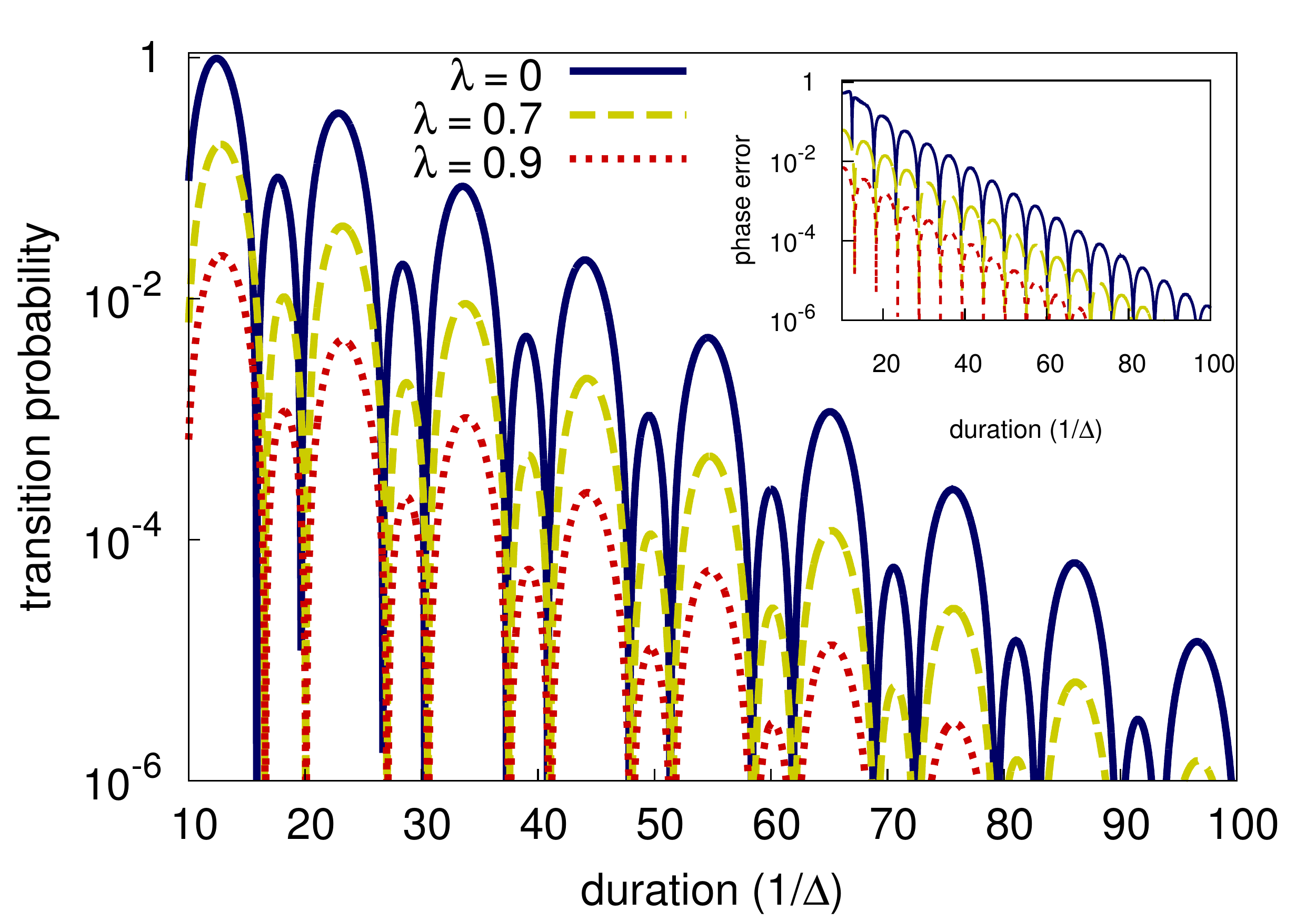}
\caption{Diabatic errors {\em vs} duration of braiding protocol for the transition probability out of the degenerate subspace of the initial state. The inset shows the phase error relative to the nonabelian Berry phase. For both quantities, curves are shown in the absence of counterdiabatic terms [$\lambda=0$ in Eq.~(\ref{lambda})] and with counterdiabatic terms with 10\% ($\lambda=0.9$) and 30\% ($\lambda=0.7$) relative error. There would be no diabatic error if the counterdiabatic errors were implemented exactly. } 
\label{fig3}
\end{figure}

Exponentially small transition rates can be realized by choosing $\Delta_1(t)=\Delta\sin^2\varphi(t)$ and $\Delta_3(t)=\Delta\cos^2\varphi(t)$. Now the gap assumes a minimum during the protocol as in the familiar Landau-Zener process. For the numerical calculation presented in Fig.~\ref{fig3} we have chosen $\varphi(t)$ to have a smooth derivative. The diabatic transition rate is indeed exponential in the protocol duration which is somewhat conterintuitive as the transition rate actually {\em decreases} relative to the previously discussed protocols although the gap is {\em smaller}. The phase error also exhibits exponential scaling as shown in the inset of Fig.\ \ref{fig3}.

An exact implementation of the counterdiabatic terms fully corrects for these errors. As can be seen from Fig.\ \ref{fig3}, a suppression by two orders of magnitude merely requires an implementation which is accurate at the 10\% level. Even a very rough implementation at the 30\% level still substantially reduces the errors. More generally, we find that the relative error scales approximately as $(1-\lambda)^2$ with the accuracy of the implementation of $H_1$. It is also worth noting that the approximate counterdiabatic terms suppress the diabatic error, but do not modify its scaling with protocol duration. 

{\em Conclusions.---}In summary, we have generalized the concept of shortcuts to adiabaticity to nonabelian Berry phases and showed how this can in principle be used to implement nonabelian braiding operations exactly in a finite time. Such protocols can substantially improve the accuracy of braiding operations performed in a finite time interval. It is interesting to note that our scheme bears some resemblance with the concept of quasi-adiabatic continuity for topological phases  \cite{hastings}.

In this work we have focused on a simple model of nonabelian braiding which excludes the quasiparticle continuum. The current protocols are therefore useful whenever there is a separation of scales between the finite-energy subgap states and the magnitude of the topological gap. Including the quasiparticle continuum is an interesting problem for future research. It should also be interesting to extend the current considerations for Majorana zero modes to more exotic nonabelian quasiparticles. 

{\em Acknowledgments.---}We acknowledge useful discussions with Piet Brouwer and Aris Alexandradinata. This work was funded by the Packard Foundation and the Institute for Quantum Information and Matter, an NSF Physics Frontiers Center with support of the Gordon and Betty Moore Foundation through Grant GBMF1250. We are also grateful for financial support by the Virtual Institute ``New states of matter and their excitations" as well as DFG Schwerpunkte 1666 ``Topological insulator" and 1459 ``Graphene."

\clearpage
\onecolumngrid

\section{Supplementary material}

\subsection{Derivation of Eq.\ (14) in the main text}

Here we derive general expressions for the counterdiabatic correction for non-interacting systems in terms of the (instantaneous) eigenmodes $\gamma_{n,\alpha}$ of the original (time-dependent) Hamiltonian in a Majorana representation,
\begin{align}
 H_0=i\sum_{n\alpha}\epsilon_n\gamma_{n,2\alpha-1}\gamma_{n,2\alpha}.
\end{align}
Note that both the eigen-Majoranas $\gamma_{n,\alpha}$ and the eigenvalues $\epsilon_n$ are time dependent. Degeneracies of the many-body spectrum can arise when one or more of the eigenvalues $\epsilon_n$ vanish or when some nonzero $\epsilon_n$ is degenerate, independent of time. The various Majorana operators associated with each single-particle eigenvalue $\epsilon_n$ are labeled by the index $\alpha$. If the single-particle eigenvalue $\epsilon_n$ is $N$-fold degenerate, $\alpha$ takes on $2N$ different values, $\alpha=1,\ldots,2N$.   

A direct derivation of the counterdiabatic terms based on the general Eq.\ (12) in the main text is cumbersome. Here, we choose to proceed as follows. The counterdiabatic terms suppress transitions out of the degenerate subspace but leave the dynamics within the degenerate subspace as governed by the nonabelian Berry connection of the original time evolution unchanged. Thus, we can determine the counterdiabatic terms $H_1$ uniquely from the following constraints:
\begin{itemize}
\item[(a)] $H_1$ has no matrix elements which act within the degenerate eigenspaces of $H_0$. This ensures that $H_1$ affects only transitions between states with different energies.
\item[(b)] The time evolution of the $\gamma_{n,\alpha}$ with respect to the full shortcut Hamiltonian $H=H_0+H_1$ does not include any transitions between different degenerate subspaces. 
\end{itemize}
To implement these constraints, we note that the time evolution of the Majorana operators is governed by the Heisenberg equation of motion
\begin{equation}
  \frac{d\gamma_{n,\alpha}}{dt}=\frac{\partial\gamma_{n,\alpha}}{\partial t}+i[H_0+H_1,\gamma_{n,\alpha}],
\label{HEOM}
\end{equation}
where the first term on the right-hand side accounts for the explicit time dependence of the Majorana operators. This term allows for the expansion
\begin{equation}
   \frac{\partial\gamma_{n,\alpha}}{\partial t} = \sum_{m,\beta} C_{nm}^{\alpha\beta} \gamma_{m,\beta}. 
\end{equation}
Here, the coefficients $C_{nm}^{\alpha\beta}$ can in principle be expressed in terms of the instantaneous eigenfunctions and their time derivatives. It turns out, however, that we do not need these explicit expressions for the present purpose. To proceed, we also write the counterdiabatic terms in the general form
\begin{equation}
    H_1=i\sum_{n,m}\sum_{\alpha,\beta} h^{\alpha,\beta}_{n,m} \gamma_{n,\alpha}\gamma_{m,\beta}.
\end{equation}
Thus, it is our goal to derive the coefficients $h^{\alpha,\beta}_{n,m}$ which satisfy the antisymmetry relation $h^{\alpha,\beta}_{n,m}=-h^{\beta,\alpha}_{m,n}$.

Implementing the contraints (a) and (b), we demand that the $\gamma_{n,\alpha}$ satisfy the time evolution
\begin{equation}
  \frac{d\gamma_{n,\alpha}}{dt}= \sum_{\beta} C_{nn}^{\alpha\beta}\gamma_{n,\beta} + i[H_0,\gamma_{n,\alpha}].
\label{desire}
\end{equation}
Here, the first term of the right-hand side contains only those terms of $ \frac{\partial\gamma_{n,\alpha}}{\partial t}$ that belong to the same subspace $n$. All terms in  $ \frac{\partial\gamma_{n,\alpha}}{\partial t}$ which belong to different subspaces must be cancelled by the counterdiabatic terms $H_1$. 

The desired shortcut time evolution in Eq.\ (\ref{desire}) satisfies 
\begin{equation}
   \left\{\frac{d\gamma_{n,\alpha}}{d t},\gamma_{m,\beta}\right\}=0
\end{equation}
for $m\neq n$. Inserting the Heisenberg equation of motion (\ref{HEOM}) into this condition, we obtain 
\begin{equation}
      4h^{\alpha\beta}_{nm}-4h^{\beta\alpha}_{mn}+\left\{\frac{\partial\gamma_{n,\alpha}}{\partial t},\gamma_{m,\beta}\right\}=0
\end{equation}
for $m\neq n$. Using the antisymmetry property of the $h^{\alpha\beta}_{nm}$ yields
\begin{align}
   H_1=-\frac{i}{8}\sum_{\substack{n,m\\ (n\neq m)}}\sum_{\alpha\beta}\left\{\frac{\partial\gamma_{n,\alpha}}{\partial t},\gamma_{m, \beta}\right\}\gamma_{n,\alpha}\gamma_{m,\beta}.
\end{align}
Finally, we write this as 
\begin{align}
   H_1=\frac{i}{8}\sum_{n,m}\sum_{\alpha\beta}\gamma_{m,\beta}\left\{\gamma_{m, \beta},\frac{\partial\gamma_{n,\alpha}}{\partial t}\right\}\gamma_{n,\alpha}-\frac{i}{8}\sum_{n}\sum_{\alpha\beta}\gamma_{n,\beta}\left\{\gamma_{n, \beta},\frac{\partial\gamma_{n,\alpha}}{\partial t}\right\}\gamma_{n,\alpha}.
\end{align}
and use the relation
\begin{equation}
  \sum_{m}\sum_{\beta}\gamma_{m,\beta}\left\{\gamma_{m,\beta},\frac{\partial\gamma_{n,\alpha}}{\partial      
     t}\right\} = 2\frac{\partial\gamma_{n,\alpha}}{\partial  t}
\end{equation}
to obtain Eq.~(14) of the main text.

In the following, we derive the counterdiabatic terms for the braiding procedure given in Eq.\ (17) of the main text using this general result as well as a more basic approach starting with Eq.\ (12). 

\subsection{Derivation of Eq.\ (17) using the general Majorana counterdiabatic terms}\label{toughluck}

Here, we derive Eq.\ (17) using $H_1$ in the Majorana operator representation as given in Eq.~(14). The braiding Hamiltonian in Eq.~(16) is
\begin{equation}
H_0=ih_\Delta(t)\gamma_0\gamma_\Delta(t).\label{h0_app}
\end{equation}
The system comprises a mode of energy $h_\Delta(t)$ associated with the two Majorana operators $\gamma_0$ and $\gamma_\Delta(t)$ as defined in the main text. In addition there are two Majoranas $\gamma_A(t)$ and $\gamma_B(t)$ which remain uncoupled by Eq.~(\ref{h0_app}) and form a zero-energy mode. Using the identity
\begin{align}
 \{ \dot{\gamma}_\alpha,\gamma_\beta \}=- \{ \dot{\gamma}_\beta,\gamma_\alpha\}
\end{align}
(with the shorthand $\dot \gamma=\frac{\partial\gamma}{\partial t}$) we can write the time derivatives (suppressing time arguments) in the most generic form as 
\begin{align}
 \dot{\gamma}_A&=\eta_{1}\gamma_B+\eta_{2}\gamma_\Delta,\nonumber\\
 \dot{\gamma}_B&=-\eta_{1}\gamma_A+\eta_{3}\gamma_\Delta,\nonumber\\
 \dot{\gamma}_\Delta&=-\eta_{2}\gamma_A-\eta_{3}\gamma_B.
\end{align}
with real coefficients $\eta_i$. The first term in $H_1$ can be written as
\begin{align}
 \sum_{n\alpha} \dot{\gamma}_{n,\alpha}\gamma_{n,\alpha}=\dot{\gamma}_A\gamma_A+\dot{\gamma}_B\gamma_B+ \dot{\gamma}_\Delta\gamma_\Delta=2\dot{\gamma}_\Delta \gamma_\Delta+2\eta_{1}\gamma_B\gamma_A
\end{align}
The second contribution to $H_1$ subtracts all terms within a degenerate subspace and thus eliminates the term $\sim \gamma_B\gamma_A$. Thus we obtain
\begin{align}
H_1=\frac{i}{2}\dot{\gamma}_\Delta\gamma_\Delta=\frac{i} {2h_\Delta^2}\sum\limits_{\alpha\beta} \dot\Delta_{\alpha}\Delta_{\beta}\gamma_{\alpha}\gamma_{\beta}
\end{align}
as given in Eq.\ (17). 

\subsection{Derivation of Eq.\ (17) using the spin construction}

We can alternatively derive Eq.\ (17) using the general formulation of the counterdiabatic terms in Eq.~(12). To this end, we introduce conventional fermionic operators through
\begin{equation}
   c_1 = \frac{1}{2} (\gamma_1 - i\gamma_2) \,\,\,\, ; \,\,\,\, c_2 = \frac{1}{2}(\gamma_0 -i \gamma_3).
\end{equation}
Using the inverse relations 
\begin{equation}
  \gamma_1 = c_1 + c_1^\dagger  \,\, ; \,\, \gamma_2 = i(c_1 - c_1^\dagger) \,\, ; \,\, \gamma_3 = i(c_2 - c_2^\dagger) \,\, ; \,\, 
  \gamma_0 = c_2 + c_2^\dagger  ,
\end{equation}
we can write $H_0$ in terms of $c_1$ and $c_2$
\begin{eqnarray}
  H_0 &=& i\sum_{j=1}^{3} \Delta_j \gamma_0\gamma_j \nonumber\\
    &=& i\Delta_1 (c_2^\dagger c_1 - c_1^\dagger c_2 + c_2c_1 - c_1^\dagger c_2^\dagger) 
     - \Delta_2  (c_2^\dagger c_1 + c_1^\dagger c_2 + c_2c_1 + c_1^\dagger c_2^\dagger) 
     -\Delta_3 (2c_2^\dagger c_2 -1).
\end{eqnarray}
Specifically, we write the Hamiltonian in the basis $\{|00\rangle,|11\rangle,|10\rangle,|01\rangle \} $, where the basis states are defined as 
\begin{equation}
   |11\rangle = c_1^\dagger c_2^\dagger |00\rangle \,\, , \,\, |10\rangle = c_1^\dagger |00\rangle \,\, , \,\, |01\rangle = c_2^\dagger |00\rangle
\end{equation} 
with $c_1 |00\rangle = c_2 |00\rangle =0$. This yields
\begin{equation}
  H_0 = \left( \begin{array} {cccc} \Delta_3 & i\Delta_1 - \Delta_2 & 0 & 0 \cr
                           -i\Delta_1 - \Delta_2 & -\Delta_3 & 0 & 0 \cr
                            0 & 0 & \Delta_3 & -i\Delta_1 - \Delta_2 \cr
                             0 & 0 & i\Delta_1 - \Delta_2 & - \Delta_3 \cr \end{array}\right) .
\label{44ham}
\end{equation}
The block-diagonal structure originates from the conservation of fermion-number parity. In fact, it is easy to show that the Hamiltonian $H$ commutes with the parity operator
\begin{equation}
    P = \gamma_0\gamma_1\gamma_2\gamma_3.
\end{equation}
The top-left block $H_{\rm even}= \Delta_3\tau_z - \Delta_1 \tau_y - \Delta_2 \tau_x$ corresponds to even fermion parity, while the bottom-right block $H_{\rm odd}=\Delta_3\tau_z + \Delta_1 \tau_y - \Delta_2 \tau_x$ has odd fermion parity. Here we have defined Pauli matrices $\tau_i$ within the even and odd subspaces. If we also define Pauli matrices $\pi_j$ in the even-odd subspace, then we can write  
\begin{equation}
  H_0 = \Delta_3\tau_z - \Delta_1 \tau_y \pi_z - \Delta_2 \tau_x
  \label{h0taupi}
\end{equation}
for the overall Hamiltonian $H$. Expressing $H_{\rm even}$ and $H_{\rm odd}$ in terms of Pauli matrices makes it obvious that these Hamiltonians take the form of a spin Hamiltonian in magnetic fields $B_{\rm even} = (-\Delta_2,-\Delta_1,\Delta_3)$ and $B_{\rm odd} = (-\Delta_2,\Delta_1,\Delta_3)$, respectively. The degeneracy due to the presence of the Majorana modes implies that the two subspaces have the same eigenvalues. At the same time, the spectrum for each subspace by itself is non-degenerate. 

In order to evaluate the counterdiabatic terms, it is useful to eliminate the time derivatives of the states from Eq.\ (12). To achieve this, we first multiply the first term on the right-hand side of Eq.\ (12) by ${\bf 1} = \sum_m\sum_\beta |\psi^m_\beta\rangle\langle \psi^m_\beta|$ from the left and obtain
\begin{equation}
   H_1 = i\sum_{m\neq n}\sum_{\alpha\beta} |\psi^m_\beta\rangle \langle \psi^m_\beta|\partial_t \psi^n_\alpha\rangle\langle \psi_\alpha^n|.  
\label{h11}
\end{equation}
Taking the time derivative of Eq.\ (5) and multiplying from the left by $\langle \psi_\beta^m|$, one finds
\begin{equation}
   \langle \psi^m_\beta|\partial_t \psi^n_\alpha\rangle = \frac{\langle \psi_\beta^m|\partial_t H_0|\psi_\alpha^n\rangle}{E_n-E_m}
\end{equation}
for $n\neq m$. Inserting this into Eq.\ (\ref{h11}) yields
\begin{equation}
   H_1 = i\sum_{m\neq n}\sum_{\alpha\beta} |\psi^m_\beta\rangle \frac{\langle \psi_\beta^m|\partial_t H_0|\psi_\alpha^n\rangle}{E_n-E_m} \langle \psi_\alpha^n|.  
\label{h111}
\end{equation}
Using this expression, the counterdiabatic terms $H_1$ can be conveniently derived. 

To do so, we temporarily perform rotations within the even and odd subspaces such that $H_0(t)$ in Eq.\ (\ref{h0taupi}) involves only the $\tau_z$ term. Then, the eigenstates in the even and odd subspaces are simply the ``spin-up" and the ``spin-down" states. Using
\begin{equation}
   \partial_t H_0 = \dot\Delta_3\tau_z - \dot\Delta_1 \tau_y \pi_z - \dot\Delta_2 \tau_x,
\end{equation}
we then find that 
\begin{equation}
  H_1 = \frac{1}{2(\Delta_1^2+\Delta_2^2+\Delta_3^2)} \left\{ (\Delta_3\dot\Delta_1 - \Delta_1\dot\Delta_3) \tau_x \pi_z 
   + (\Delta_2\dot\Delta_3 - \Delta_3\dot\Delta_2) \tau_y - (\Delta_1\dot\Delta_2 - \Delta_2\dot\Delta_1) \tau_z \pi_z\right\}.
\end{equation}
This can be readily expressed in terms of the original Majorana operators. Indeed, we have the identities
\begin{eqnarray}
  i\Delta_{12} \gamma_1\gamma_2 &=& -\Delta_{12}(2c_1^\dagger c_1 -1) \nonumber \\
  i\Delta_{13} \gamma_1\gamma_3 &=& \Delta_{13}(c_2 c_1 + c_1^\dagger c_2^\dagger - c_2^\dagger c_1 - c_1^\dagger c_2) \\
  i\Delta_{23} \gamma_2\gamma_3 &=& i\Delta_{23}(c_2 c_1 - c_1^\dagger c_2^\dagger + c_1^\dagger c_2 - c_2^\dagger c_1), \nonumber
\end{eqnarray}
or, in the basis specified above, 
\begin{eqnarray}
  i\Delta_{12} \gamma_1\gamma_2 &=& \Delta_{12}\tau_z\pi_z \nonumber \\
  i\Delta_{13} \gamma_1\gamma_3 &=& \Delta_{13} \tau_x\pi_z \\
  i\Delta_{23} \gamma_2\gamma_3 &=& -\Delta_{23}\tau_y. \nonumber
\end{eqnarray}
Thus, we finally find 
\begin{equation}
   H_1 = \frac{i}{2(\Delta_1^2+\Delta_2^2+\Delta_3^2)} \left\{(\Delta_2\dot\Delta_1 - \Delta_1\dot\Delta_2)\gamma_1\gamma_2 + (\Delta_3\dot\Delta_1 - \Delta_1\dot\Delta_3)\gamma_1\gamma_3
   + (\Delta_3\dot\Delta_2 - \Delta_2\dot\Delta_3)\gamma_2\gamma_3 \right\}
\end{equation}
in terms of the original Majorana operators.

\begin{figure}[t]
\includegraphics[width=.48\textwidth]{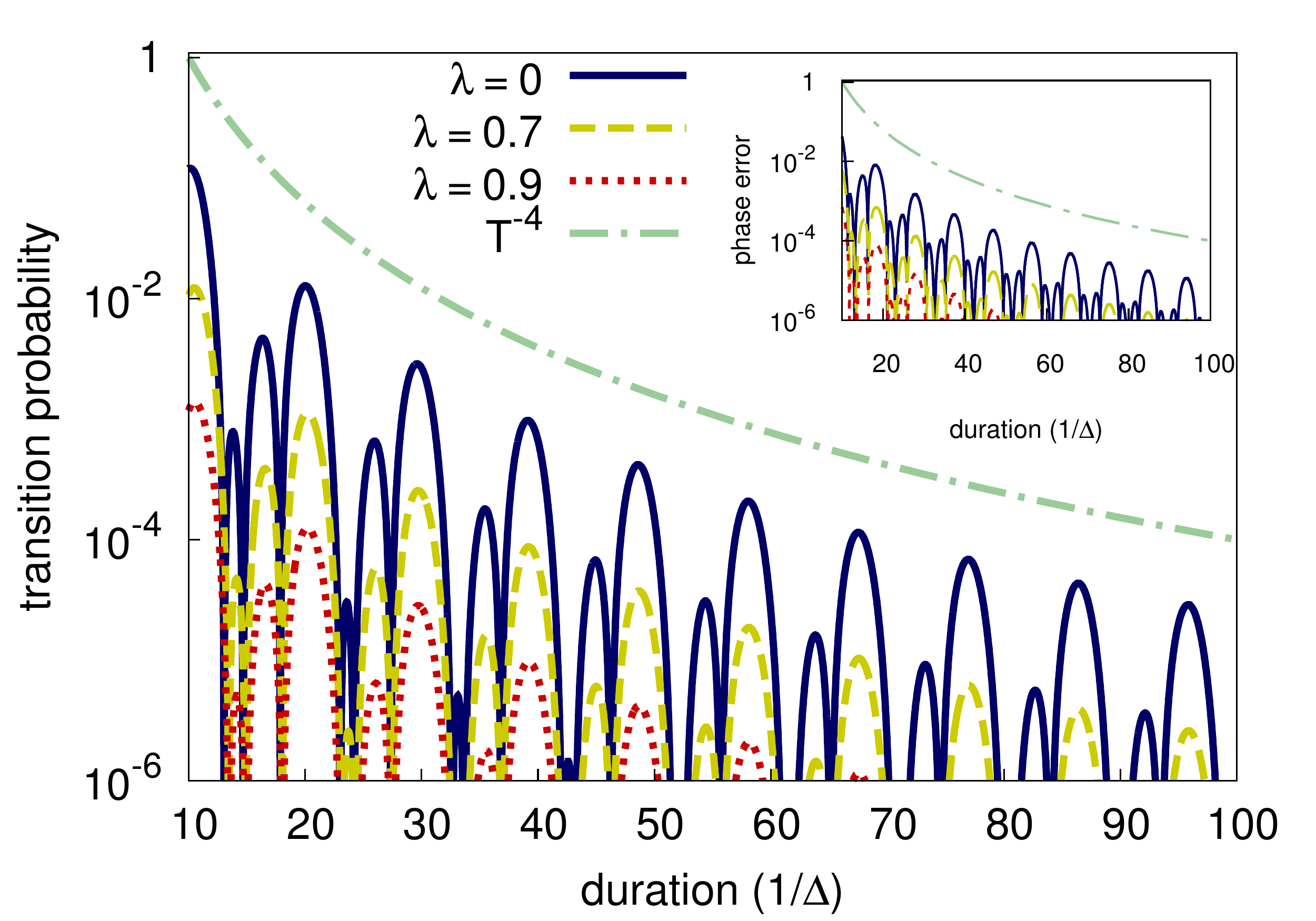}
\caption{Diabatic errors {\em vs} duration of the braiding protocol defined by $\Delta_1(t)=\Delta\sin\varphi(t)$ and $\Delta_3(t)=\Delta\cos\varphi(t)$ for the transition probability out of the degenerate subspace of the initial state. The inset shows the phase error relative to the nonabelian Berry phase. The excitation gap remains unchanged during the entire braiding protocol, which results in a power-law dependence on the duration. The function $\varphi(t)$ is chosen to have zero derivative at both endpoints. For both transition probability and phase error, curves are shown in the absence of counterdiabatic terms and with counterdiabatic terms with 10\% and 30\% relative error. There would be no diabatic error if the counterdiabatic errors were implemented exactly.} 
\label{fig5}
\end{figure}

\subsection{Numerical calculation of the robustness}

In this section we provide details of the numerical calculations. For completeness we also include numerical results for the transition probability and Berry phase errors of the non-exponential protocols mentioned in the main text. Due to the conservation of fermion-number parity the nonabelian Berry phase takes the form $\exp(i\gamma\tau_z)$, where $\tau_z$ is a Pauli matrix in parity space. Performing the braiding protocol adiabatically yields $\gamma=\pi/4$.
For finite durations we numerically compute the Berry phase as $\gamma=\arg[\langle\Psi_e(T)\Psi_e(0)\rangle/\langle\Psi_o(T)\Psi_o(0)\rangle]/2$, where $\Psi_{e/o}(t)$ denotes the ground state wavefunction at time $t$ with even (odd) parity. The diabatic phase error is $|\gamma-\pi/4|$.

We first consider the protocol with the basic step $\Delta_1(t)=\Delta\sin\varphi(t)$ and $\Delta_3(t)=\Delta\cos\varphi(t)$. When $\varphi(t)$ has zero derivative at both endpoints ($\varphi=0$ for $t=0$ and $\varphi=\pi/2$ for $t=T/3$), the transition probability scales as $T^{-4}$ with the protocol duration $T$. This is shown in Fig.\ \ref{fig5}. Specifically, we have chosen $\varphi(t) =(\pi/2)[ 3(3t/T)^2-2(3t/T)^3]$ between 0 and $t=T/3$ for this calculation (as well as for the one in the main text). When the derivative jumps at either (or both) end points (as for the simplest choice $\varphi(t) = 3\pi t/2T$), we find the errors to decay even more slowly, namely as $T^{-2}$. Note that both the transition probability out of the degenerate subspace and the phase error of the topological qubit scale in the same manner with $T$, see inset of Fig.\ \ref{fig5}. 

\begin{figure}[t]
\includegraphics[width=.48\textwidth]{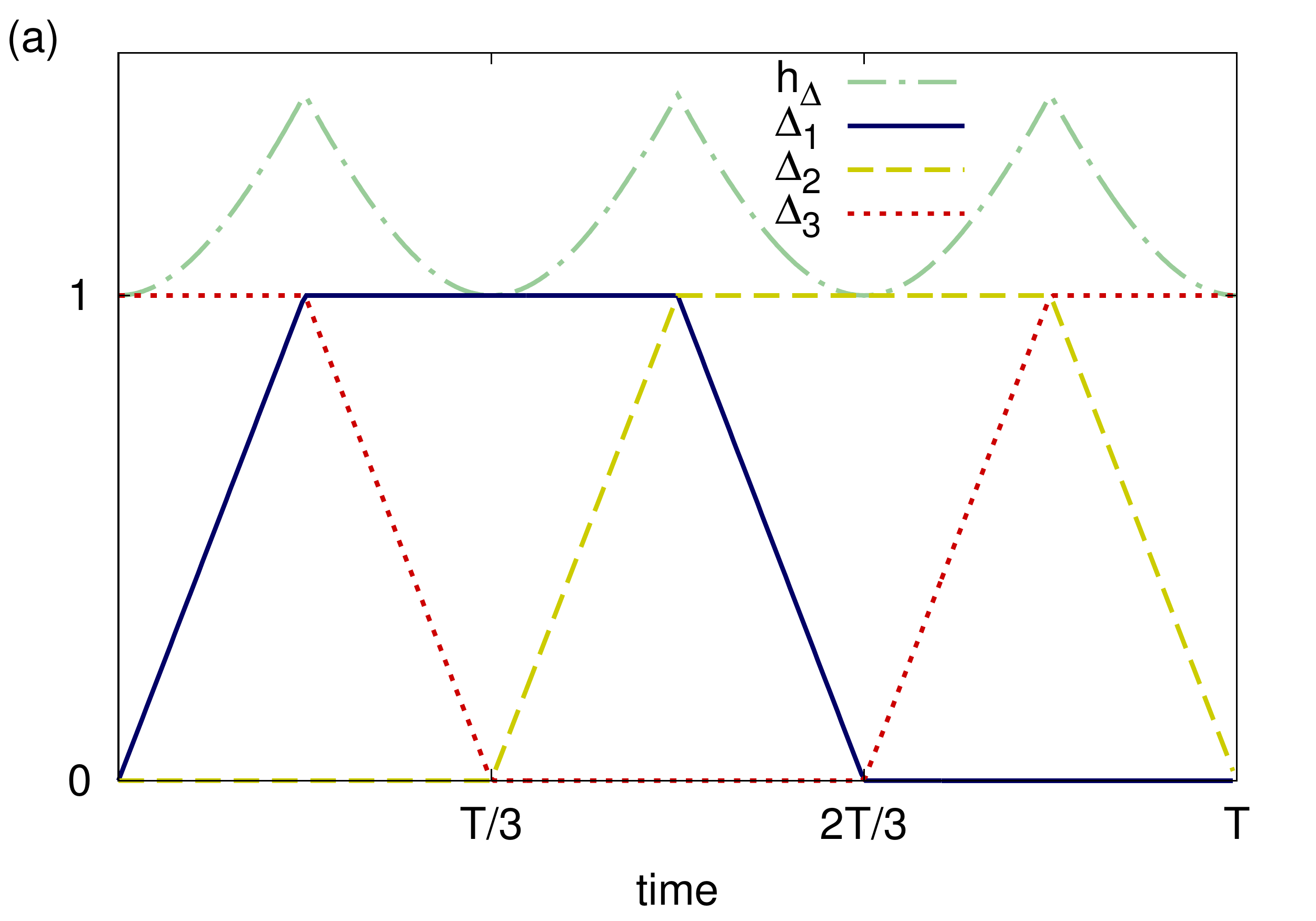}
\includegraphics[width=.48\textwidth]{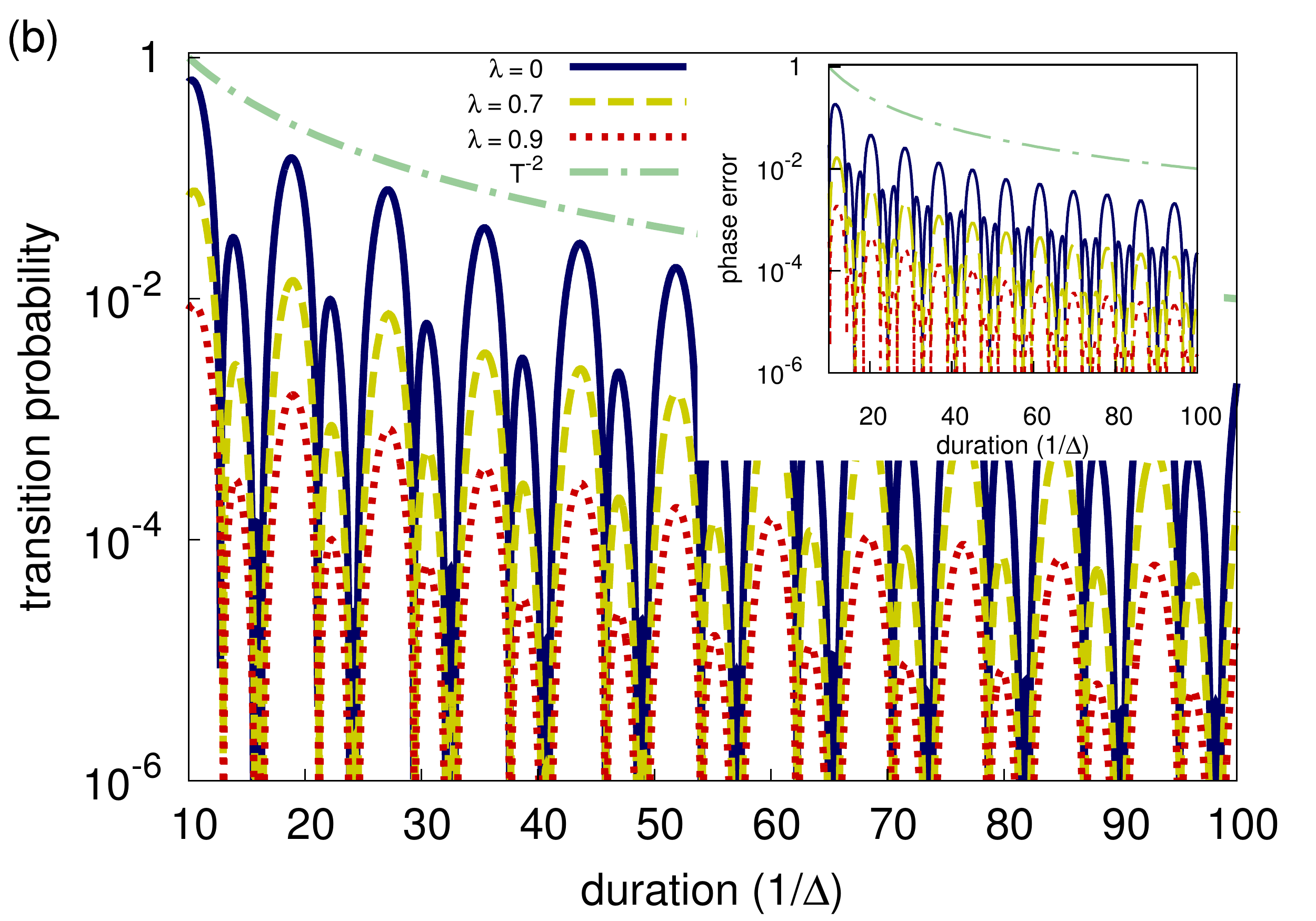}
\caption{(a) Couplings $\Delta_j$ and gap $h_\Delta$ during the braiding process for the protocol in Ref.\ [30]. (b) Diabatic errors {\em vs} duration of braiding protocol. Similar to the protocol in Fig.~\ref{fig5} the error has a power-law dependence on the duration. The derivative of $\phi(t)$ jumps at the end points and therefore the error scales as $T^{-2}$. 
} 
\label{fig6}
\end{figure}

Interestingly, the same dependences are found for the protocol given in Ref.\ [30] and displayed in Fig.~\ref{fig6}(a). The initial step of the braiding operation is effected by increasing $\Delta_1$ first at constant $\Delta_3$. The latter is reduced to zero only subsequently. In this protocol, the gap increases and takes on a maximum halfway through this basic step. Nevertheless, the diabatic errors still vary as a power law of $T$. Fig.\ \ref{fig6}(b) shows corresponding numerical results. Here we chose a linear protocol, $\varphi(t) = 3\pi t/2T$, in which the derivatives of $\varphi(t)$ do not vanish at the end points.

\end{document}